\documentclass[letterpaper,12pt]{article}
\usepackage[margin=1in]{geometry}

\usepackage{breakurl}  
\usepackage[activate={true,nocompatibility},final,tracking=false]{microtype}

\usepackage[english]{babel}
\usepackage{csquotes}

\usepackage[numbers]{natbib}

\usepackage[hidelinks]{hyperref}
\begin{document}
\frenchspacing
\begin{center}
   {\LARGE Enabling External Scrutiny of AI Systems\\\vspace{.3em}
   with Privacy-Enhancing Technologies}
   \vspace{1em}
    
   {\href{mailto:kb1520@georgetown.edu}{Kendrea Beers} and \href{mailto:helen.toner@georgetown.edu}{Helen Toner}\vspace{0.7em}
   
   Georgetown University's Center for Security and Emerging Technology (CSET)}
   \vspace{1em}
\end{center}


\begin{abstract}
    
This article describes how technical infrastructure developed by the
nonprofit OpenMined enables external scrutiny of AI systems without
compromising sensitive information.

Independent external scrutiny of AI systems provides crucial
transparency into AI development, so it should be an integral component
of any approach to AI governance. In practice, external researchers have
struggled to gain access to AI systems because of AI companies'
legitimate concerns about security, privacy, and intellectual property.

But now, privacy-enhancing technologies (PETs) have reached a new level
of maturity: end-to-end technical infrastructure developed by OpenMined
combines several PETs into various setups that enable privacy-preserving
audits of AI systems. We showcase two case studies where this
infrastructure has been deployed in real-world governance scenarios:
``Understanding Social Media Recommendation Algorithms with the
Christchurch Call'' and ``Evaluating Frontier Models with the UK AI
Safety Institute.'' We describe types of scrutiny of AI systems that
could be facilitated by current setups and OpenMined's proposed future
setups.

We conclude that these innovative approaches deserve further exploration
and support from the AI governance community. 
Interested policymakers can focus on empowering researchers on a legal level.
\end{abstract}

\section{Introduction}\label{introduction}

From the recommender system underlying Facebook's newsfeed to the large
language model powering ChatGPT, large-scale AI systems are now
operating at global scale and shaping the lives of billions of users.
The extraordinary reach of these systems has raised widespread concerns
about their risks, from personalized misinformation to biased
decision-making algorithms to novel cyberattacks against critical infrastructure.

Although experts disagree on the extent and relative severity of
different AI threats, one thing they almost all agree on is the need for independent external scrutiny of consequential AI
systems  \citep{raji_outsider_2022, shevlane_model_2023, criddle_uk_2023, ntia_comment}. Given the scale and influence of these systems, there is a
clear public interest in better understanding how they work and how they
impact society.

In practice, however, efforts to facilitate scrutiny by external
researchers have faced challenges. The owners of AI systems have been
reluctant to grant access to third-party researchers due to legitimate
concerns about compromising the data privacy of their users, the
security of their systems, or the value of their intellectual property.
Although research teams have managed to carry out studies, current approaches
are difficult to scale; they typically require lengthy legal
agreements, access to a secure physical facility, and/or significant
limitations on the questions that can be asked  \citep{audit}. Routine external
scrutiny of consequential AI systems will require scalable technical
infrastructure and streamlined bureaucratic procedures.

Recent work that operationalizes privacy-enhancing technologies (PETs)
has brought this goal within reach: the nonprofit
\textbf{OpenMined has built technical AI governance infrastructure that
enables privacy-preserving audits of AI systems.}
In the past year, this
infrastructure has been deployed in collaboration with a slew of
organizations including the Christchurch Call (a multi-stakeholder
coalition combating extremist content online), Anthropic (a frontier AI
company), and the United Kingdom AI Safety Institute. \textbf{These
demonstrations show the way toward a future where external researchers
can scrutinize AI systems with much greater ease, without compromising
sensitive information.}

\section{The Technology}\label{the-technology}

Over the past few decades, researchers in a variety of disciplines have
developed techniques for privacy-preserving study of sensitive
information. Recently, these fields---ranging from cryptography to
distributed systems to machine learning---have embraced the umbrella
term ``privacy-enhancing technologies.'' Only even more recently have
these technologies been incorporated into end-to-end infrastructure that
can flexibly adapt to different use cases.

OpenMined's end-to-end technical infrastructure draws on several
well-established privacy-enhancing technologies, combining them into
various setups that enable various kinds of privacy-preserving research.
These technologies, including secure enclaves  \citep{secure_enclave}, secure
multi-party computation  \citep{smpc}, zero-knowledge
proofs  \citep{zk}, federated
learning  \citep{federated_learning}, and differential
privacy  \citep{differential_privacy}, enable users to verify their trust in a system's relevant
privacy and security properties.

The core component of OpenMined's technical infrastructure is an
open-source software library called
PySyft  \citep{pysyft} that works with the
popular programming language Python. The infrastructure supports the
following core workflow: \textbf{a researcher remotely proposes
questions to a model owner, the model owner approves the researchers'
questions, and then the researcher receives the answers to the questions
without learning anything else about the model owner's proprietary
systems.} OpenMined has proposed a range
of specific setups, all based on this core workflow, that can vary
depending on budget, how much the model owner and the researcher trust
each other, and what types of questions the researcher wants to ask  \citep{audit}.

\section{Case Study 1: Understanding Social Media Recommendation
Algorithms with the Christchurch
Call}\label{case-study-1-understanding-social-media-recommendation-algorithms-with-the-christchurch-call}

In the aftermath of the 2019 Christchurch terrorist shootings in New
Zealand, a community of more than 120 governments, online service
providers, and civil society organizations came together as the
Christchurch Call to tackle the effects of extremist content online. In
2022, the Christchurch Call launched the Initiative
on Algorithmic Outcomes (CCIAO) to develop privacy-enhancing software
infrastructure that can enable independent researchers to scrutinize
algorithmic systems (including AI systems) without compromising user
privacy, security, or intellectual property  \citep{christchurch_2022}. Phase 1 of the CCIAO in
2023 showcased a collaboration between OpenMined and French video
platform Dailymotion (as well as similar collaborations with Twitter and
LinkedIn), which marked the first-ever use of an integrated
privacy-preserving access tool to enable external research on social
media platforms  \citep{christchurch_phase1}.

\textbf{This pilot program provided a proof of concept that external
researchers can leverage private assets (in this case, video impression
data) to investigate societal impacts of algorithms without seeing the
data.} This precluded the need for extensive legal review, making the
process far less onerous for both the researchers and the company.

Phase 1 of the CCIAO demonstrated OpenMined's simplest setup.
Dailymotion uses an AI algorithm to recommend videos, so the dataset
under scrutiny connected information like a video's ID, title, and tags
to the video's level of ``recommendedness'' by the algorithm. This
dataset provided the information necessary to determine whether the
algorithm is amplifying certain types of content.

First, Dailymotion used OpenMined's toolkit to upload their proprietary
data to a secure server. Dailymotion also provided some mock data to the
external researchers to help the researchers develop their code. Next,
the researchers proposed a topic to the Dailymotion team: investigating
the relationship between videos' suggestiveness scores and how much they
are amplified by Dailymotion's algorithms. The researchers also provided
code and an explanation of precisely how they would investigate the
research question. Finally, Dailymotion reviewed the proposal and
allowed the researchers' code to run. The researchers found that
Dailymotion's recommendation algorithms do not disproportionately
amplify highly suggestive content.

\section{Case Study 2: Evaluating Frontier Models with the UK AI
Safety
Institute}\label{case-study-2-evaluating-frontier-models-with-the-uk-ai-safety-institute}

More recently, OpenMined partnered with the United Kingdom AI Safety
Institute (UK AISI) and the frontier AI company Anthropic to run a trial
of a method for conducting safety evaluations of frontier AI models
while keeping sensitive information private  \citep{enclaves_for_evals}. Anthropic provided the AI
model, and UK AISI provided evaluation code that made use of a dataset
of questions and answers about biology. The setup was designed to enable
mutual secrecy, meaning that the contents of the biology dataset
remained private to UK AISI while the AI model weights remained private
to Anthropic. Anthropic did have access to the evaluation code that UK
AISI used to feed inputs from the dataset into the model. For the
purposes of this exercise, both the AI model
(GPT-2  \citep{gpt2})
and the biological dataset
(CAMEL
Biology  \citep{camel_bio}) were public assets, but the exercise served as a trial run for
a higher-stakes scenario in which both the model and the evaluation data
would need to be private.

\textbf{The exercise successfully demonstrated that a government entity
and an AI company can negotiate and enforce shared governance over
evaluations of frontier models.} Specifically, the focus was on ensuring
that each party (1) received all the information it needed, and (2)
could approve or deny all of the necessary computations during the
evaluation. Scaling up this setup to accommodate the largest frontier
models will require significant engineering work, but the foundational
research challenges have already been addressed.

OpenMined developed a more sophisticated technical setup to enable this
exercise. To simultaneously prevent the external evaluator from
accessing the AI model and prevent the model owner from accessing secret
information (e.g.~about bioweapons), OpenMined used a secure enclave to
host both the AI model and the biological dataset during the evaluation.
The secure enclave produced certificates proving that it only ran the
evaluation code that both Anthropic and UK AISI approved. Once the AI
model and the dataset were deployed to the secure enclave, running the
evaluation took under two hours. Compared to the setup in Case Study 1,
this setup provided two key additional features. First, the evaluator
was able to propose their own inputs to the model, rather than relying
on existing data from the model owner. Second, the evaluator was able to
bring a sensitive dataset that remained hidden from the model owner.

\section{Future Directions for Privacy-Preserving Scrutiny of AI
Systems}\label{future-directions-for-privacy-preserving-scrutiny-of-ai-systems}

\textbf{OpenMined's existing technical setups have the flexibility to
apply to a variety of paradigms of external scrutiny of AI systems.} The
simple setup for Case Study 1 enables researchers to analyze user logs
that reveal trends in any AI system's real behavior when deployed---for
example, whether a recommender system has a partisan lean, or whether a
chatbot gives toxic responses. Using secure enclave technology to allow
researchers to bring their own datasets, as shown in Case Study 2, does
not just enable government entities to leverage datasets that are kept
secret for national security reasons: it could also make it more
feasible for auditors to operate as private businesses by using audit
prompts as intellectual property, and prevent AI companies from
``teaching to the test'' on audits by training their models on audit
prompts.

\textbf{OpenMined's proposals for more sophisticated setups open up ways
to scrutinize AI systems with stronger privacy and verification
guarantees.} OpenMined is currently developing features that will allow
researchers to keep their code private in addition to their data; this
will empower researchers to conduct more sophisticated research without
as much oversight by model owners. Future setups could also include
verification mechanisms to prevent model owners from gaming audits by
deleting a model's prediction logs or switching out an audited model for
an unaudited one.

It is worth noting that OpenMined's technical infrastructure can
facilitate not just AI research, but also privacy-preserving research on
digital systems more broadly. For example, OpenMined recently kicked off
a beta
program with Reddit that will enable researchers to responsibly access
Reddit data in bulk  \citep{reddit}. OpenMined is also supporting the national
statistical agencies of the US, Canada (StatCan), and Italy (IStat) in
conducting joint privacy-preserving
analyses of their data  \citep{mitchell_new_2024}.

\section{Conclusion}\label{conclusion}

Now that privacy-enhancing technologies have reached a new level of
maturity, AI model owners can no longer use privacy, security, and IP as
conclusive excuses for refusing access for external researchers. This
means that interested \textbf{policymakers can now focus on empowering
researchers on a legal level}. One aspect of empowering researchers is
promoting legal safe harbor for
AI auditors, i.e.~protecting them from account suspensions and legal
reprisal  \citep{safe_harbor}. A second aspect is providing legal backing for research
projects that AI model owners may wish to decline. Although the above
case studies demonstrate that some AI model owners are willing to
approve external research projects, many model owners may not be so
willing; recall that, in deployments so far, OpenMined's workflow has
given model owners the right to approve or deny all research questions.
Legislation could serve as a backstop by requiring AI model owners to
comply with important research nonetheless. For example, the proposed Platform
Accountability and Transparency Act (PATA) would require social media
companies to make internal information available to researchers whose
projects have been vetted by the National Science Foundation for this
purpose  \citep{pata}.

In summary, external scrutiny of AI systems provides crucial
transparency into AI development, so it should be an integral component
of any approach to AI governance. Historically, external researchers
have struggled to gain access to AI systems because of AI companies'
valid concerns about security, privacy, and intellectual property. But
now, trustworthy privacy-preserving technical solutions for external
scrutiny of AI systems have succeeded in real-world governance
scenarios. These innovative approaches deserve further exploration and
support from the AI governance community.

\pagebreak
{\small
\bibliographystyle{plainnat}
\bibliography{sample}

\begin{thebibliography}{20}
\providecommand{\natexlab}[1]{#1}
\providecommand{\url}[1]{\texttt{#1}}
\expandafter\ifx\csname urlstyle\endcsname\relax
  \providecommand{\doi}[1]{doi: #1}\else
  \providecommand{\doi}{doi: \begingroup \urlstyle{rm}\Url}\fi

\bibitem[Chen et~al.(2024)Chen, Bandy, Buckley, and Bhatia]{christchurch_phase1}
Jiahao Chen, Jack Bandy, Dave Buckley, and Ruchi Bhatia.
\newblock {AI} transparency in practice, October 2024.
\newblock URL \url{https://www.christchurchcall.org/safe-secure-private-research-finds-third-parties-can-audit-online-algorithms/}.

\bibitem[{Christchurch Call}(2022)]{christchurch_2022}
{Christchurch Call}.
\newblock Christchurch {Call} initiative on algorithmic outcomes, September 2022.
\newblock URL \url{https://www.christchurchcall.org/christchurch-call-initiative-on-algorithmic-outcomes/}.

\bibitem[Criddle(2023)]{criddle_uk_2023}
Cristina Criddle.
\newblock {UK} pushes for greater access to {AI}’s inner workings to assess risks.
\newblock \emph{Financial Times}, September 2023.
\newblock URL \url{www.ft.com/content/4427689c-b98e-483e-bde3-2d82808a266f}.

\bibitem[{Hugging Face}(2019)]{gpt2}
{Hugging Face}.
\newblock {OpenAI} {GPT2}, 2019.
\newblock URL \url{https://huggingface.co/docs/transformers/model_doc/gpt2}.

\bibitem[KeyserSosa(2024)]{reddit}
KeyserSosa.
\newblock Our plans for researchers on {Reddit}, May 2024.
\newblock URL \url{www.reddit.com/r/reddit4researchers/comments/1co0mqa/our_plans_for_researchers_on_reddit/}.

\bibitem[Li et~al.(2023)Li, Hammoud, Itani, Khizbullin, and Ghanem]{camel_bio}
Guohao Li, Hasan Abed Al~Kader Hammoud, Hani Itani, Dmitrii Khizbullin, and Bernard Ghanem.
\newblock Camel: communicative agents for “mind” exploration of large scale language model society, May 2023.
\newblock URL \url{https://huggingface.co/datasets/camel-ai/biology}.

\bibitem[Longpre et~al.(2024)Longpre, Kapoor, Klyman, Ramaswami, Bommasani, Blili-Hamelin, Huang, Skowron, Yong, Kotha, Zeng, Shi, Yang, Southen, Robey, Chao, Yang, Jia, Kang, Pentland, Narayanan, Liang, and Henderson]{safe_harbor}
Shayne Longpre, Sayash Kapoor, Kevin Klyman, Ashwin Ramaswami, Rishi Bommasani, Borhane Blili-Hamelin, Yangsibo Huang, Aviya Skowron, Zheng-Xin Yong, Suhas Kotha, Yi~Zeng, Weiyan Shi, Xianjun Yang, Reid Southen, Alexander Robey, Patrick Chao, Diyi Yang, Ruoxi Jia, Daniel Kang, Sandy Pentland, Arvind Narayanan, Percy Liang, and Peter Henderson.
\newblock A safe harbor for {AI} evaluation and red teaming, March 2024.
\newblock URL \url{https://arxiv.org/abs/2403.04893}.

\bibitem[Lopardo(2020)]{federated_learning}
Antonio Lopardo.
\newblock What is federated learning?, May 2020.
\newblock URL \url{https://blog.openmined.org/what-is-federated-learning/}.

\bibitem[Lopardo et~al.(2020)Lopardo, Benaissa, and Ryffel]{smpc}
Antonio Lopardo, Ayoub Benaissa, and Théo Ryffel.
\newblock What is secure multi-party computation?, May 2020.
\newblock URL \url{https://blog.openmined.org/what-is-secure-multi-party-computation/}.

\bibitem[Maniam et~al.(2023)Maniam, Nelson, Garfinkel, Christian, Ho, Chou, Toner, Raji, Solaiman, Phillips, Perset, Aidinoff, Botvinick, Salganik, Chowdhury, Bowman, Krier, Barocas, Friedler, Ifayemi, and Isaac]{ntia_comment}
Aaron Maniam, Alondra Nelson, Ben Garfinkel, Brian Christian, Daniel~E. Ho, Dorothy Chou, Helen Toner, Inioluwa~Deborah Raji, Irene Solaiman, James~W. Phillips, Karine Perset, Marc Aidinoff, Matthew Botvinick, Matthew~J. Salganik, Rumman Chowdhury, Samuel~R. Bowman, Sebastien Krier, Solon Barocas, Sorelle Friedler, Stephanie Ifayemi, and William~S. Isaac.
\newblock Comment of the {AI} policy and governance working group on the {NTIA AI} accountability policy request for comment docket {NTIA-230407-0093}, 2023.
\newblock URL \url{www.ias.edu/sites/default/files/AI%20Policy%20and%20Governance%20Working%20Group%20NTIA%20Comment.pdf}.

\bibitem[Mitchell(2024)]{mitchell_new_2024}
Curtis Mitchell.
\newblock A new model for international, privacy-preserving data science.
\newblock \emph{2024 USENIX Conference on Privacy Engineering Practice and Respect}, June 2024.
\newblock URL \url{www.usenix.net/conference/pepr24/presentation/mitchell}.

\bibitem[{Office of {U.S.} {Senator} {Christopher} {Coons} of {Delaware}}(2021)]{pata}
{Office of {U.S.} {Senator} {Christopher} {Coons} of {Delaware}}.
\newblock {Coons}, {Portman}, {Klobuchar} announce legislation to ensure transparency at social media platforms, December 2021.
\newblock URL \url{www.coons.senate.gov/news/press-releases/coons-portman-klobuchar-announce-legislation-to-ensure-transparency-at-social-media-platforms}.

\bibitem[OpenMined(2020)]{differential_privacy}
OpenMined.
\newblock Introduction to differential privacy, 2020.
\newblock URL \url{https://openmined.github.io/PyDP/introduction.html}.

\bibitem[OpenMined(2021)]{pysyft}
OpenMined.
\newblock {PySyft}'s documentation, 2021.
\newblock URL \url{https://openmined.github.io/PySyft/}.

\bibitem[Raji et~al.(2022)Raji, Xu, Honigsberg, and Ho]{raji_outsider_2022}
Inioluwa~Deborah Raji, Peggy Xu, Colleen Honigsberg, and Daniel~E. Ho.
\newblock Outsider oversight: designing a third party audit ecosystem for {AI} governance, June 2022.
\newblock URL \url{https://arxiv.org/abs/2206.04737}.

\bibitem[S. and Hall(2020)]{zk}
Srishilesh~P. S. and Adam~J. Hall.
\newblock What are zero knowledge proofs?, October 2020.
\newblock URL \url{https://blog.openmined.org/zero-knowledge-proof/}.

\bibitem[Shevlane et~al.(2023)Shevlane, Farquhar, Garfinkel, Phuong, Whittlestone, Leung, Kokotajlo, Marchal, Anderljung, Kolt, Ho, Siddarth, Avin, Hawkins, Kim, Gabriel, Bolina, Clark, Bengio, Christiano, and Dafoe]{shevlane_model_2023}
Toby Shevlane, Sebastian Farquhar, Ben Garfinkel, Mary Phuong, Jess Whittlestone, Jade Leung, Daniel Kokotajlo, Nahema Marchal, Markus Anderljung, Noam Kolt, Lewis Ho, Divya Siddarth, Shahar Avin, Will Hawkins, Been Kim, Iason Gabriel, Vijay Bolina, Jack Clark, Yoshua Bengio, Paul Christiano, and Allan Dafoe.
\newblock Model evaluation for extreme risks, September 2023.
\newblock URL \url{https://arxiv.org/abs/2305.15324}.

\bibitem[Strahm(2023)]{secure_enclave}
Lacey Strahm.
\newblock Response to the {National Telecommunications and Information Administration (NTIA)} request for comment {(RFC)} on dual use foundation artificial intelligence {(AI)} models with widely available model weights, 2023.
\newblock URL \url{https://downloads.regulations.gov/NTIA-2023-0009-0334/attachment_1.pdf}.

\bibitem[Trask et~al.(2023)Trask, Sukumar, Kalliokoski, Farkas, Ezenwaka, Popa, Mitchell, Hrebenach, Muraru, Junior, Bejan, Mishra, Ngong, Bandy, Stahl, Cardonnet, Trask, Trask, Nguyen, Dang, van~der Veen, Eng, Strahm, Ayre, Jay, Lytvyn, Kyemenu-Sarsah, Chung, Smith, S, Falcon, Gupta, Gabriel, Milea, Thoraldson, Porto, Cebere, Gorana, and Reza]{audit}
Andrew Trask, Akshay Sukumar, Antti Kalliokoski, Bennett Farkas, Callis Ezenwaka, Carmen Popa, Curtis Mitchell, Dylan Hrebenach, George-Cristian Muraru, Ionesio Junior, Irina Bejan, Ishan Mishra, Ivoline Ngong, Jack Bandy, Jess Stahl, Julian Cardonnet, Kellye Trask, Kellye Trask, Khoa Nguyen, Kien Dang, Koen van~der Veen, Kyoko Eng, Lacey Strahm, Laura Ayre, Madhava Jay, Oleksandr Lytvyn, Osam Kyemenu-Sarsah, Peter Chung, Peter Smith, Rasswanth S, Ronnie Falcon, Shubham Gupta, Stephen Gabriel, Teo Milea, Theresa Thoraldson, Thiago Porto, Tudor Cebere, Yash Gorana, and Zarreen Reza.
\newblock How to audit an {AI} model owned by someone else ({P}art 1), June 2023.
\newblock URL \url{https://blog.openmined.org/ai-audit-part-1/}.

\bibitem[Trask et~al.(2024)Trask, Yesilyurt, Farkas, Ezenwaka, Popa, Buckley, van~der Wel, Mosconi, Han, Junior, Bejan, Mishra, Nguyen, van~der Veen, Eng, Strahm, Graham, Jay, Simtinica, Kyemenu-Sarsah, Smith, S, Falcon, Wagh, Mandala, Gupta, Gabriel, Ramkumar, Ahmed, Milea, Maggio, Gorana, and Reza]{enclaves_for_evals}
Andrew Trask, Aziz~Berkay Yesilyurt, Bennett Farkas, Callis Ezenwaka, Carmen Popa, Dave Buckley, Eelco van~der Wel, Francesco Mosconi, Grace Han, Ionesio Junior, Irina Bejan, Ishan Mishra, Khoa Nguyen, Koen van~der Veen, Kyoko Eng, Lacey Strahm, Logan Graham, Madhava Jay, Matei Simtinica, Osam Kyemenu-Sarsah, Peter Smith, Rasswanth S, Ronnie Falcon, Sameer Wagh, Sandeep Mandala, Shubham Gupta, Stephen Gabriel, Subha Ramkumar, Tauquir Ahmed, Teo Milea, Valerio Maggio, Yash Gorana, and Zarreen Reza.
\newblock Secure enclaves for {AI} evaluation, November 2024.
\newblock URL \url{https://blog.openmined.org/secure-enclaves-for-ai-evaluation/}.

\end{thebibliography}
}
\end{document}